\documentclass{jfm}

\usepackage{graphicx}
\usepackage{natbib}
\usepackage{amsmath}
\usepackage{amssymb}
\usepackage{hyperref}
\usepackage{nicefrac}
\usepackage{xcolor}

\title[Turbulent RB convection described by projected dynamics in phase space]{Turbulent Rayleigh--B{\'e}nard convection described by projected dynamics in phase space}

\author[J.~L{\"u}lff , M.~Wilczek, R.~J.~A.~M.~Stevens, R.~Friedrich, D.~Lohse]%
{Johannes L{\"u}lff$^1$\thanks{Email address for correspondence: johannes.luelff@uni-muenster.de}, Michael Wilczek$^{2}$, Richard J.~A.~M.~Stevens$^{3,4}$, Rudolf Friedrich$^1$\thanks{deceased}, Detlef Lohse$^4$}

\affiliation{$^1$Institute for Theoretical Physics, University of M{\"u}nster, Wilhelm-Klemm-Str.~9, D-48149 M{\"u}nster, Germany\\[\affilskip]
	$^2$Max Planck Institute for Dynamics and Self-Organization, D-37077 G{\"o}ttingen, Germany\\[\affilskip]
	$^3$Department of Mechanical Engineering and Center for Environmental and Applied Fluid Mechanics, The Johns Hopkins University, 3400 North Charles Street, Baltimore, Maryland 21218, USA\\[\affilskip]
	$^4$Physics of Fluids Group, Department of Science and Technology, Mesa+ Institute, and J.~M.~Burgers Center for Fluid Dynamics, University of Twente, 7500 AE Enschede, The Netherlands}

\begin{document}

\maketitle

\begin{abstract}
	Rayleigh--B{\'e}nard convection, i.e.~the flow of a fluid between two parallel plates that is driven by a temperature gradient, is an idealised setup to study thermal convection.
	Of special interest are the statistics of the turbulent temperature field, which we are investigating and comparing for three different geometries, namely convection with periodic horizontal boundary conditions in three and two dimensions as well as convection in a cylindrical vessel, in order to work out similarities and differences.
	To this end, we derive an exact evolution equation for the temperature probability density function (PDF).
	Unclosed terms are expressed as conditional averages of velocities and heat diffusion, which are estimated from direct numerical simulations.
	This framework lets us identify the average behaviour of a fluid particle by revealing the mean evolution of fluid of different temperatures in different parts of the convection cell.
	We connect the statistics to the dynamics of Rayleigh--B{\'e}nard convection, giving deeper insights into the temperature statistics and transport mechanisms.
	We find that the average behaviour is described by closed cycles in phase space that reconstruct the typical Rayleigh--B{\'e}nard cycle of fluid heating up at the bottom, rising up to the top plate, cooling down and falling down again.
	The detailed behaviour shows subtle differences between the three cases.
\end{abstract}

\renewcommand{\vec}[1]{\boldsymbol{#1}}
\renewcommand{\Pr}{\mbox{\textit{Pr}}}
\newcommand{\Ra}{\mbox{\textit{Ra}}}
\newcommand{\bcdot}{\boldsymbol{\cdot}}

\section{Introduction}
	In Rayleigh--B{\'e}nard convection, a fluid enclosed between two horizontal plates is heated from below and cooled from above, which induces a flow and thereby enhanced heat transport between the plates.
	This simple setup is the benchmark system to study thermal convection, which is important in nature and technical applications.
	Prominent examples include convection in the oceans and the atmosphere or plate tectonics in the mantle of the earth.
	Depending on the control parameters, the Rayleigh--B{\'e}nard system displays a variety of different patterns and flow regimes, ranging from laminar to highly turbulent flows.

	Apart from special cases like laminar convection, an analytical solution does not exist as turbulent flows are remarkably hard to handle analytically.
	Because of their importance, a deeper understanding of turbulent convective flows is still desired, despite the inability to solve the basic equations analytically.
	To achieve this, many different approaches have been pursued.
	The heat transport as function of the control parameters is of particular interest and is well described by the Grossmann--Lohse theory \citep{grossmann00jfm,grossmann01prl,ahlers09rmp,grossmann11pof,grossmann12pof,petschel13prl,stevens13jfm}.
	There have also been studies on the turbulence properties of Rayleigh--B{\'e}nard convection, by e.g.~characterising the statistics of temperature readings of thermal probes in the convection cell \citep{yakhot89prl,ching93prl,ching04prl,shang08prl} or by examining the heat transport mechanisms and the large-scale circulation by Eulerian \citep{bailoncuba10jfm,petschel11pre, poel11pre, ahlers12njp} or Lagrangian \citep{gasteuil07prl,schumacher09pre} approaches.
	An overview of recent progress on Rayleigh--B{\'e}nard convection can be found in \citet{ahlers09rmp,lohse10rfm,chilla12epj}.

	In this paper we will describe Rayleigh--B{\'e}nard convection with the full single-point temperature statistics using the temperature probability density function (PDF).
	This in turn gives us information about the dynamics of the convecting fluid.
	To this end, we will derive an evolution equation for the temperature PDF, feed in numerical data to complete our ansatz and obtain through the statistics a description of the mean dynamics of fluid particles that travel around in the convection cell.
	A similar method has been used to describe the statistics in homogeneous isotropic turbulence before, see \citet{wilczek09pre,wilczek11jfm,friedrich12crp}.
	In this paper, we will generalise and extend the work presented by \citet{luelff11njp}, where we first introduced the aforementioned method to Rayleigh--B{\'e}nard convection.

	We start by deriving the framework in the most general form.
	Since Rayleigh--B{\'e}nard setups usually contain a number of symmetries that can be utilised to simplify the problem, we will apply our framework to three different showcases of three- and two-dimensional convection with homogeneous horizontal directions (i.e., periodic boundaries) and three-dimensional convection in a cylinder.
	All three cases have different statistical symmetries and show slightly different dynamics.
	The differences between two- and three-dimensional convection and also between fixed side walls and periodic horizontal boundaries are discussed, for example, in \citet{poel13jfm,poel14pre}.
	We will use the PDF methods presented here to further work out similarities and differences between these three cases and give a comprehensive description of the statistics and the dynamics of Rayleigh--B{\'e}nard convection.

	Since the derivation of our framework utilises the basic equations of Rayleigh--B{\'e}nard convection, it can be considered as an ansatz from first principles.
	The basic equations that govern Rayleigh--B{\'e}nard convection are the Oberbeck--Boussinesq equations \citep{oberbeck79apc,boussinesq03book} for the velocity $\vec{u}(\vec{x},t)$ and temperature field $T(\vec{x},t)$:
	\begin{subequations}\label{eq:boussinesq}
		\begin{align}
			\frac{\partial}{\partial t} \vec{u} + \vec{u}\bcdot \vec{\nabla} \vec{u} &= -\vec{\nabla} p + \Pr \Delta \vec{u} + \Pr \Ra T \vec{e}_z \label{eq:boussinesq1}\\
			\vec{\nabla}\bcdot\vec{u}&=0 \label{eq:boussinesq1.5}\\
			\frac{\partial}{\partial t} T + \vec{u}\bcdot \vec{\nabla} T &= \Delta T \label{eq:boussinesq2}
		\end{align}
	\end{subequations}
	Here, the equations have been non-dimensionalised by the heat diffusion time $\frac{L^2}{\kappa}$, the vertical height $L$ and the heat difference $\Delta$ between the upper and lower plate.
	This introduces the Rayleigh number $\Ra=\frac{\alpha g \Delta L^3}{\nu\kappa}$ and the Prandtl number $\Pr=\frac{\nu}{\kappa}$ as the control parameters.
	The vertical coordinate lies in the range $z\in[0,1]$ and the temperature takes values $T\in[0,1]$.
	Another control parameter that is often taken into account is the aspect ratio $\Gamma$ which indicates the lateral over the vertical extent of the system.

	The remainder of this paper is structured as follows.
	In Sec.~\ref{sec:theory} we will briefly recount our method, i.e.~derive an equation for the temperature PDF and connect it to the description of the dynamics of the system.
	This general theory is then applied to three different Rayleigh--B{\'e}nard geometries in Sec.~\ref{sec:dns_results}, namely three- and two-dimensional convection with homogeneous horizontal directions, and three-dimensional convection in a closed cylindrical vessel with $\Gamma=1$.
	Section~\ref{sec:discussion} closes the article with an interpretation and discussion of the findings.

\section{Statistical Description of Heat Transport}\label{sec:theory}
	Our idea to describe Rayleigh--B{\'e}nard convection is to start from the temperature PDF.
	Therefore we want to derive an equation that describes the temperature PDF and use it to gain insights into the dynamics of the system.
	This ansatz is generally referred to as \emph{PDF methods} \citep{pope84iaa, pope00book} or the \emph{Lundgren--Monin--Novikov hierarchy} \citep{lundgren67pof,monin67pmm,novikov68sdp}.
	We now give a short overview of this derivation; a more detailed discussion of the framework can be found in \citet{luelff11njp,wilczek09pre,wilczek11jfm,friedrich12crp}.
	Similar equations have been derived for turbulent reactive flows by \citet{pope85pec}.
	However, there the unclosed terms are modelled instead of estimated from the numerics as in our case.

\subsection{PDF Methods}
	The starting point is the definition of the temperature PDF as an ensemble average,
	\begin{equation}\label{eq:pdf}
		f(T,\vec{x},t)=\bigl\langle \delta\bigl(T(\vec{x},t)-T\bigr) \bigr\rangle,
	\end{equation}
	where the PDF $f(T,\vec{x},t)$ describes the probability to find fluid of temperature $T$ at position $\vec{x}$ and time $t$.
	Accordingly, $T$ is the sample space variable, while $T(\vec{x},t)$ is a realisation of the temperature field.
	The averaging process $\langle\cdot\rangle$ can be considered as an ensemble average; later on, ensemble averages are evaluated from the numerics by a suitable volume and time average.

	Since the definition in Eq.~\eqref{eq:pdf} includes an actual realisation $T(\vec{x},t)$ of the temperature field, it is now possible to calculate spatial and temporal derivatives of the PDF, i.e.~$\vec{\nabla} f(T,\vec{x},t)$ and $\frac{\partial}{\partial t} f(T,\vec{x},t)$.
	These derivatives contain unclosed terms in the form of conditional averages $\langle\cdot\lvert T,\vec{x},t\rangle$, where, e.g., the appearing conditionally averaged velocity $\langle\vec{u}\lvert T,\vec{x},t\rangle$ is a function of the sample space variables $T$, $\vec{x}$ and $t$ that tells us what the mean velocity is for fluid of given temperature, position and time.

	Putting the aforementioned derivatives together and rearranging them gives the desired evolution equation that describes the temperature PDF:
	\begin{subequations}\label{eq:pdfeq_general}
		\begin{align}
			\frac{\partial}{\partial t} f + \vec{\nabla}\bcdot\bigl(\langle\vec{u}\lvert T,\vec{x},t\rangle f\bigr) &= -\frac{\partial}{\partial T} \Bigl(\Bigl\langle \frac{\partial}{\partial t} T + \vec{u}\bcdot \vec{\nabla} T \Bigl\lvert T,\vec{x},t\Bigr\rangle f\Bigr) \label{eq:pdfeq_general1}\\
			&= -\frac{\partial}{\partial T} \bigl(\langle \Delta T \lvert T,\vec{x},t\rangle f\bigr)\label{eq:pdfeq_general2}
		\end{align}
	\end{subequations}
	The left-hand side of \eqref{eq:pdfeq_general1} can be seen as the convective derivative of the PDF $f(T,\vec{x},t)$, while the right-hand side of \eqref{eq:pdfeq_general1} contains the conditional average of the convective derivative of the temperature field.
	Since $T(\vec{x},t)$ is a realisation of the temperature field, it obeys the Oberbeck--Boussinesq equations, so in Eq.~\eqref{eq:pdfeq_general2} we replaced the convective derivative of the temperature field by the right-hand side of the Oberbeck--Boussinesq equation \eqref{eq:boussinesq2}.

	Above we obtained an evolution equation that links the shape of the temperature PDF to the conditionally averaged velocity $\langle\vec{u}\lvert T,\vec{x},t\rangle$ and heat diffusion $\langle\Delta T\lvert T,\vec{x},t\rangle$, which have to be supplied externally; in our case, we estimate them from simulations later on.

\subsection{Method of Characteristics}\label{sec:moc}
	The above evolution equation \eqref{eq:pdfeq_general} that determines the temperature PDF is a first-order partial differential equation.
	That means that we can apply the method of characteristics \citep{courant62book, sarra03joma} which lets us identify the average behaviour of fluid as it travels through phase space.

	In a nutshell, by applying the method of characteristics to the evolution equation, one can identify trajectories (the so-called \emph{characteristic curves} or just \emph{characteristics}) in phase space, along which the partial differential equation for the temperature PDF transforms into a set ordinary differential equations for $T$ and $\vec{x}$.
	The phase space is spanned by the variables that the temperature PDF depends upon, i.e.~$T$, $\vec{x}$ and $t$.
	The characteristics are defined by the conditional averages,
	\begin{equation}\label{eq:characteristics_general}
		\left( \begin{matrix} \dot T \\ \dot{\vec{x}} \\ \dot t \end{matrix} \right) = \left( \begin{matrix} \langle \Delta T \lvert T,\vec{x},t\rangle \\ \langle\vec{u}\lvert T,\vec{x},t\rangle \\ 1 \end{matrix} \right).
	\end{equation}
	This states that the characteristics are solutions $\bigl(T, \vec{x}, t \bigr)^\mathrm{T}$ of Eq.~\eqref{eq:characteristics_general} that follow the vector field on the right-hand side of the above equation; the vector field is regarded as the phase space velocity.
	From the last line of Eq.~\eqref{eq:characteristics_general}, $\dot{t} = 1$, it becomes clear that the parametrisation of the characteristics in phase space, i.e.~the arc length, is the same as the time of the system -- a \emph{fast} movement in phase space therefore really has to be seen in the temporal sense.

	It is now important to notice that, since the characteristics are governed by the conditionally averaged vector field, they show the average behaviour of a fluid parcel in phase space.
	In other words, the characteristics can be seen as the mean evolution of an ensemble of Lagrangian particles that share the same coordinates in phase space.
	This is what \citet[Sec.~4.5]{pope85pec} refers to as \emph{conditional particles} -- quasi-particles following the conditionally averaged vector field \eqref{eq:characteristics_general} that show the mean Lagrangian evolution and that have the same statistics as Lagrangian particles.
	By examining the conditionally averaged vector field and the resulting characteristic curves, one can investigate the mean transport properties of fluid through phase space and gain insight into the mean heat transport properties of Rayleigh--B{\'e}nard convection.
	Since the characteristics are trajectories in phase space, the framework can be seen as a quasi-Lagrangian description, but it has to be stressed that it is achieved by utilising the statistics of Eulerian fields alone.

	Along the characteristics, the partial differential equation \eqref{eq:pdfeq_general} becomes an ordinary differential equation which can be integrated. 
	Thus, the temperature PDF along a certain characteristic evolves according to
	\begin{align}\label{eq:integrated_pdf}
		f(T(t), \vec{x}(t), t) &= f(T(t_0), \vec{x}(t_0), t_0)\times\nonumber\\
		&\mathrm{exp}{\left[-\int\limits_{t_0}^t\!\mathrm{d}t'\,\Bigl(\vec{\nabla}\bcdot\langle\vec{u}\lvert T,\vec{x},t\rangle+\frac{\partial}{\partial T}\langle \Delta T \lvert  T,\vec{x},t\rangle\Bigr)_{T(t'), \vec{x}(t'), t'}\right]}.
	\end{align}
	Here, the integral is a line integral along a characteristic from $t_0$ to $t$.
	The integral kernel is the phase space divergence ${\vec{\nabla}\bcdot\langle\vec{u}\lvert T,\vec{x},t\rangle}+{\frac{\partial}{\partial T}\langle \Delta T \lvert  T,\vec{x},t\rangle}$ evaluated at the phase space position given by the characteristic for time $t'$, i.e.~$\bigl(T(t'),\vec{x}(t'),t'\bigr)^\mathrm{T}$.
	This equation tells us that the temperature PDF along the characteristic that connects the initial point $\bigl(T(t_0), \vec{x}(t_0), t_0\bigr)^\mathrm{T}$ with the point $\bigl(T(t), \vec{x}(t), t \bigr)^\mathrm{T}$ in phase space changes according to the integrated phase space divergence.
	As an alternative interpretation, Eq.~\eqref{eq:integrated_pdf} determines how the temperature PDF for $t$ is traced back to the PDF for $t_0$.
	While we will not further investigate Eq.~\eqref{eq:integrated_pdf} in the numerical results of the next section, we included it for the sake of completeness.
	
	Up to now, we have kept the description as general as possible.
	But usually, a Rayleigh--B{\'e}nard setup has a number of statistical symmetries, which simplify the problem, i.e.~the phase space dimension is reduced and the estimation of the unknown conditional averages from numerical simulation is simplified.
	In the next section, we will apply the framework that has been outlined in this section to three different Rayleigh--B{\'e}nard geometries with different symmetries and discuss the findings.

\section{Result from DNS}\label{sec:dns_results}
	In this section, we focus on three Rayleigh--B{\'e}nard geometries, i.e.~three-dimensional convection with periodic horizontal boundaries (Sec.~\ref{sec:dns_results_3d}), two-dimensional convection with periodic horizontal boundaries (Sec.~\ref{sec:dns_results_2d}), and three-dimensional convection in a cylindrical vessel with $\Gamma=1$ (Sec.~\ref{sec:dns_results_cyl}).

\subsection{Three-dimensional Convection with Periodic Horizontal Boundaries}\label{sec:dns_results_3d}
	\begin{figure}
		\centerline{\includegraphics[width=0.75\linewidth]{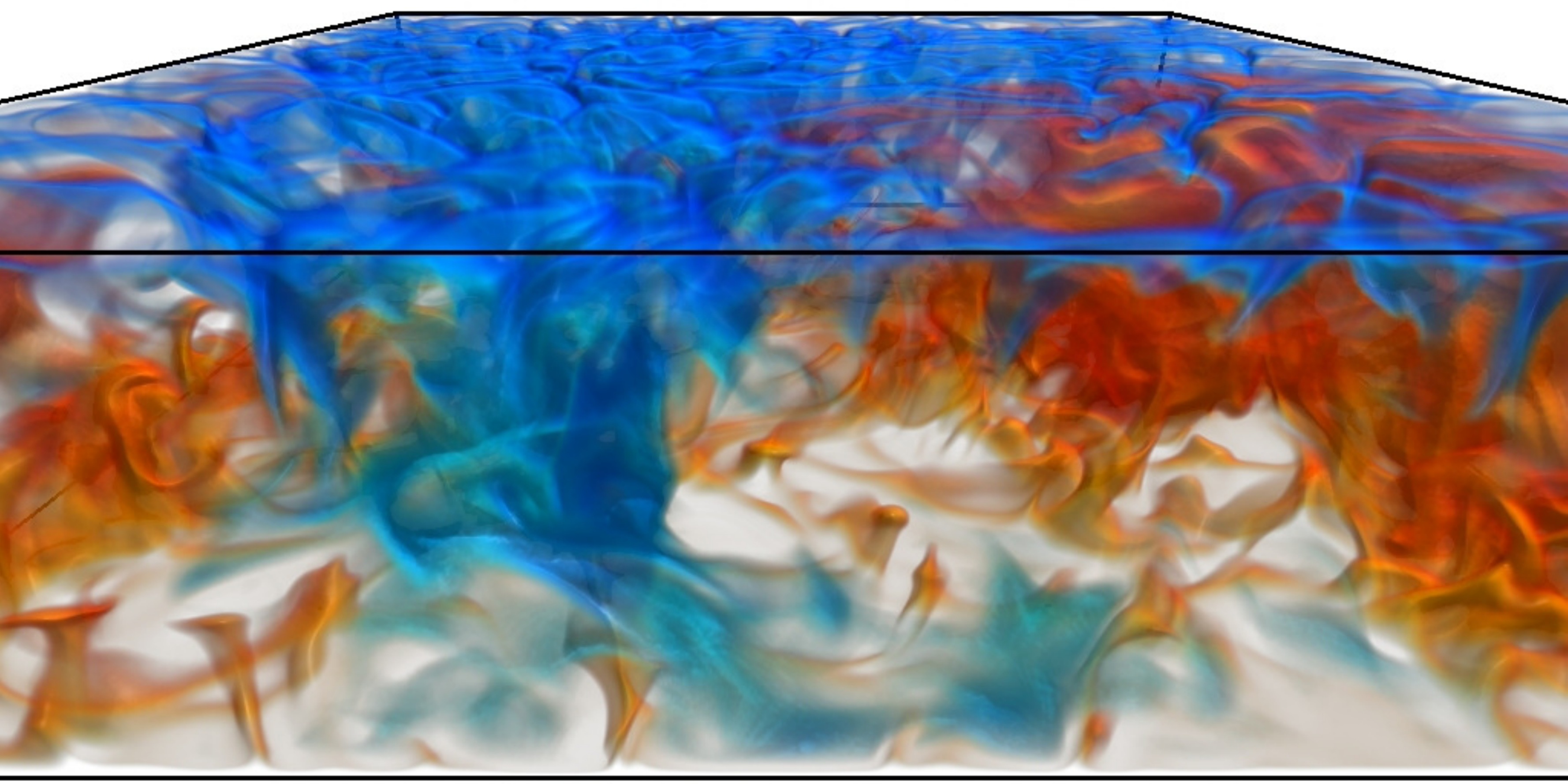}}
		\caption{\label{fig:viz_3d}
			Snapshot of the temperature field in three-dimensional Rayleigh--B{\'e}nard convection.
			Hot fluid rising up from the bottom plate is reddish, while cold fluid falling down from the top is dyed blue.
		}
	\end{figure}
	First, we consider three-dimensional convection with periodic horizontal boundaries in the statistically stationary state.
	A snapshot of the instantaneous temperature field taken from the numerics can be seen in Fig.~\ref{fig:viz_3d}.
	The parameters of the simulation are $\Ra=2.4\times10^7$, $\Pr=1$, and the aspect ratio of the periodic box is $\Gamma=4$.
	The two horizontal plates have a constant temperature and a no-slip velocity boundary condition.
	The numerical setup is a tri-periodic pseudospectral direct numerical simulation, where the boundary conditions are enforced by volume penalization methods \citep{luelff11njp,angot99num,schneider05caf,keetels07jcp}.
	The equidistant resolution in $x$, $y$ and $z$ direction is $512\times512\times128$ grid points, and in the vertical direction $5$ grid points fall into the boundary layers, according to the criteria given by \citet{shishkina10njp}.
	We calculated the statistical quantities from an ensemble consisting of $571$ snapshots, and the snapshots where taken $3.75$ free-fall time units apart.

	Statistically stationary Rayleigh--B{\'e}nard convection in this geometry is homogeneous in horizontal directions.
	This means that the statistical quantities only depend on the temperature $T$ and the vertical coordinate $z$ and not on the horizontal coordinates $x$ and $y$ or time $t$.
	Thus, the temperature PDF and the conditional averages read $f(T,z)$ and $\langle\cdot\lvert T,z\rangle$, and the phase space becomes two-dimensional.

	\begin{figure}
		\centerline{\includegraphics{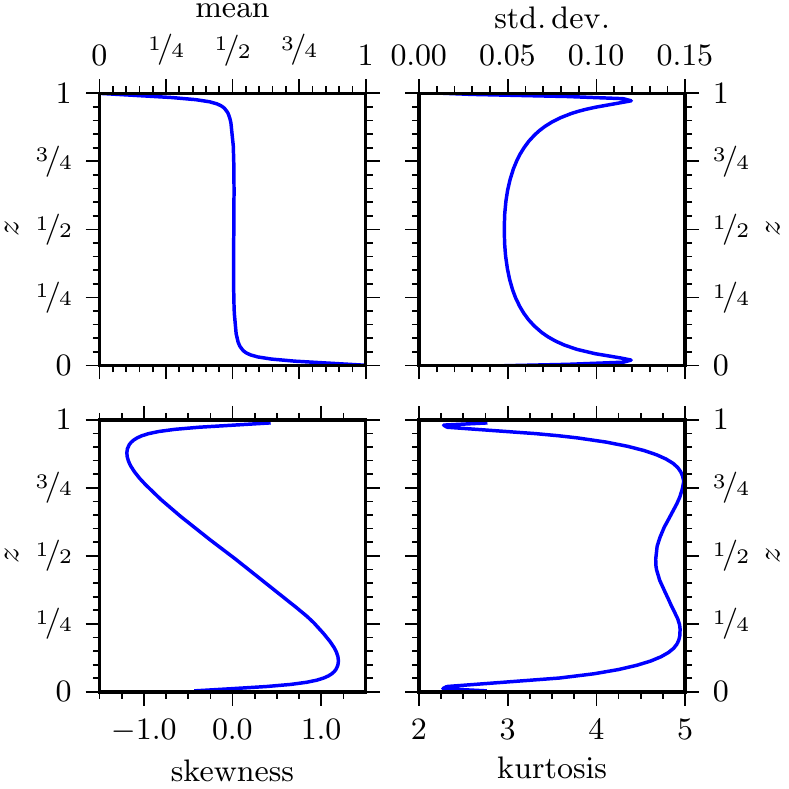}}
		\caption{\label{fig:moments_3d}
			Height-resolved profiles of the mean, standard deviation, skewness and kurtosis of the temperature distribution (row-major from upper-left panel) for three-dimensional convection with periodic horizontal boundaries.
			The skewness and kurtosis are defined as the third and fourth standardised moment.
			Gaussian distributions have skewness of $0$ and a kurtosis of $3$.
		}
	\end{figure}
	In Fig.~\ref{fig:moments_3d}, the height-resolved mean, standard deviation, skewness and kurtosis of the temperature field are shown.
	As is well known, the mean temperature is almost constant in the bulk and has a steep gradient towards the hot and cold boundaries at $z=0$ and $z=1$.
	The standard deviation takes its highest values close to the boundaries and decreases towards the centre of the bulk, indicating a temperature PDF that is broadening towards the boundaries.
	The height-resolved skewness takes its highest absolute values near the boundaries and decreases linearly as function of height in the bulk.
	This can be interpreted as hot fluid that is beginning to cool down on its way from the lower to the upper plate (and vice versa).
	The kurtosis indicates that, apart from the boundaries, the temperature PDF is more peaked and shows stronger tails than the Gaussian distribution.

	When the simplifications resulting from the statistical symmetries are incorporated into the general framework presented in Sec.~\ref{sec:theory}, Eq.~\eqref{eq:pdfeq_general} that defines the PDF becomes
	\begin{equation}\label{eq:pdfeq_3d}
		\frac{\partial}{\partial z}\bigl(\langle u_z\lvert T,z\rangle f\bigr) = -\frac{\partial}{\partial T} \bigl(\langle \Delta T \lvert T,z\rangle f\bigr),
	\end{equation}
	while the vector field \eqref{eq:characteristics_general} of the characteristics reads
	\begin{equation}\label{eq:characteristics_3d}
		\left(\begin{matrix}\dot{T} \\ \dot{z}\end{matrix}\right) = \left(\begin{matrix}\langle\Delta T\lvert T,z\rangle \\ \langle u_z\lvert T,z\rangle\end{matrix}\right).
	\end{equation}
	The PDF and the characteristic curves are therefore defined by the conditional averages of vertical velocity and heat diffusion.
	The next step is to estimate the conditional averages from the numerics while taking the statistical symmetries into account.
	Subsequently, the characteristics are obtained by integrating Eq.~\eqref{eq:characteristics_3d} for arbitrary initial conditions $\bigl(T_0, z_0\bigr)^\mathrm{T}$ in phase space.
	Obviously only initial conditions where the PDF and the conditional averages are defined, i.e.~where there have been any events at all, can be considered.

\begin{figure}
		\centerline{\includegraphics{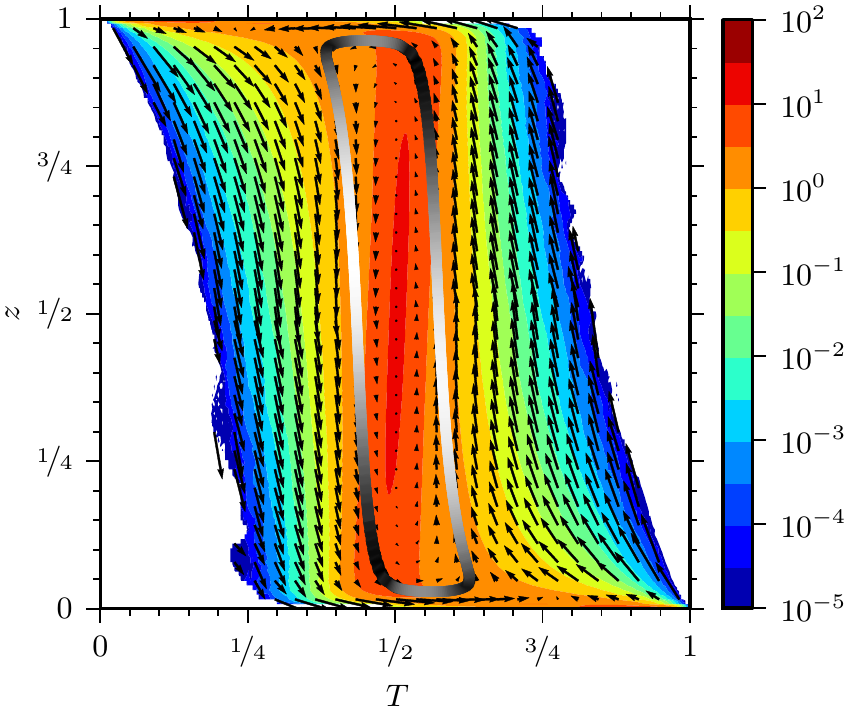}}
		\caption{\label{fig:limit_cycle_3d_pdf}
			RB cycle of the characteristics for three-dimensional convection.
			The phase space speed along the cycle shown as solid thick line is coded in black and white, i.e. the norm of the phase space velocity appearing in Eq.~\eqref{eq:characteristics_3d}.
			The colour coding in the background shows the temperature PDF $f(T,z)$.
			The arrows show the phase space velocity field, where the length of the arrows has been rescaled to arbitrary units for visualisation purposes.
			Note that around the lower-left and upper-right corners, no events were recorded (e.g., there is no fluid of temperature $T\approx1$ near the upper plate).
		}
	\end{figure}

\begin{figure}
		\centerline{\includegraphics{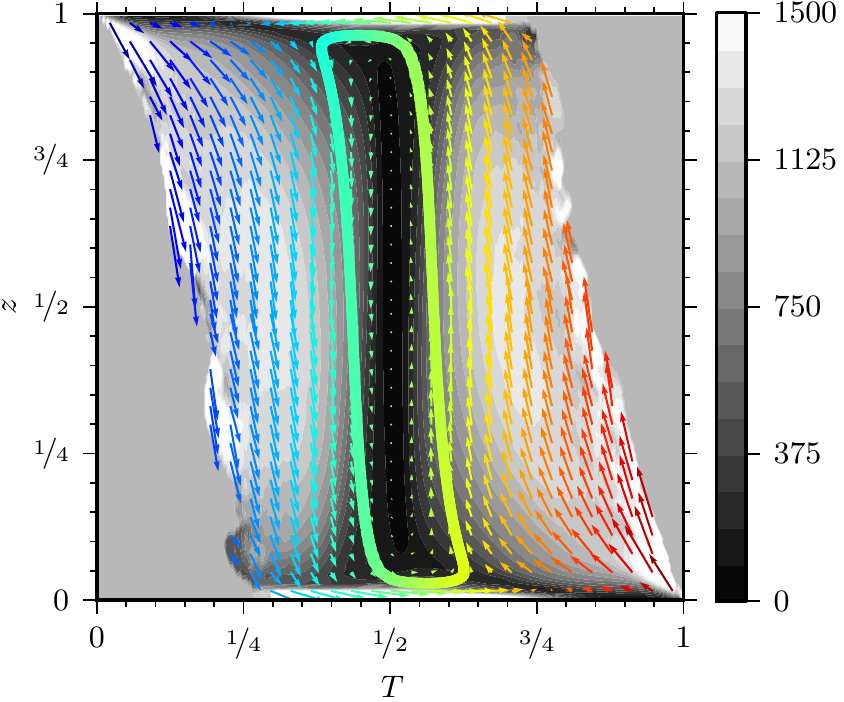}}
		\caption{\label{fig:limit_cycle_3d_vel}
			RB cycle of the characteristics for three-dimensional convection as function of height and temperature.
			The colour of the arrows (with blue corresponding to $T=0$ and red to $T=1$) indicates the temperature.
			The black and white background colour shows the phase space speed.
		}
	\end{figure}

	When integrating the characteristics for many starting positions, we observed that they tend to converge to what at first seems to be similar to a limit cycle.
	This cycle is shown in Figs.~\ref{fig:limit_cycle_3d_pdf} and~\ref{fig:limit_cycle_3d_vel}; later on, we will come back to a description of the dynamics and behaviour of Rayleigh--B{\'e}nard convection that these figures offer.
	In contrast to a limit cycle, though, one would expect that the characteristics form concentric closed curves:
	To see this, let the phase space be densely seeded by the conditional particles described in Sec.~\ref{sec:moc}.
	As the density of the conditional particles following the characteristics is proportional to the temperature PDF, and a limit cycle acts as an attractor for the conditional particles, the temperature PDF should converge towards a $\delta$-function that is non-vanishing on the cycle and zero everywhere else.
	This in turn stands in contrast to the fact that we are considering statistically stationary systems and that the temperature PDF is clearly not a $\delta$-function, and thus it follows that the characteristics cannot converge to a limit cycle but must form concentric closed curves.
	
	We find that the observed convergence is caused by the flawed conditionally averaged vector field estimated from the numerics by a binning process.
	The noise inherent to the binning violates the solenoidality of the probability flux (phase space velocity times PDF) as demanded by Eq.~\eqref{eq:pdfeq_3d}, and thus the imperfect binned vector field contains many localised sinks where the characteristics converge.
	
\begin{figure}
		\centerline{\includegraphics{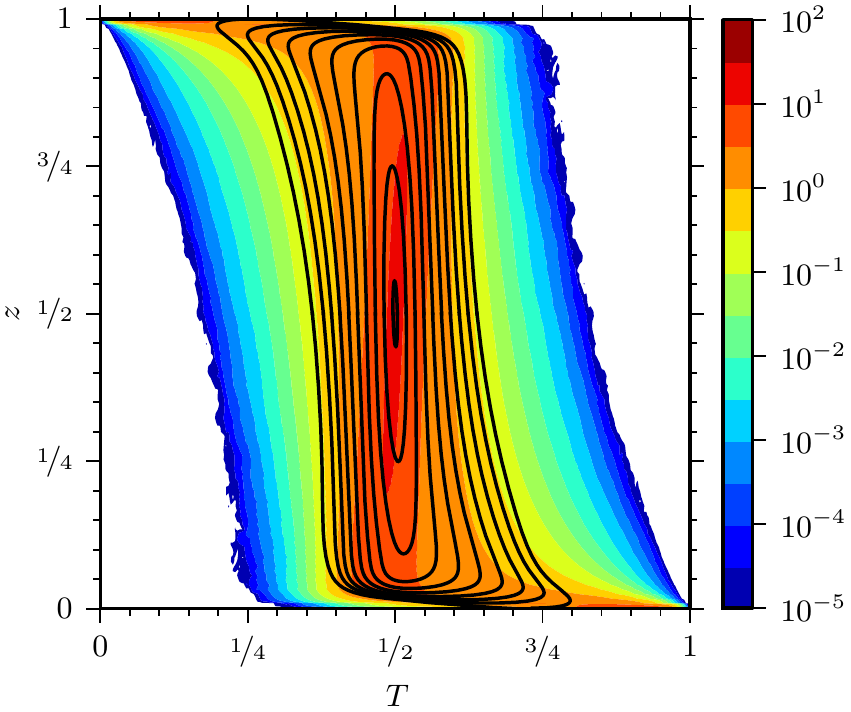}}
		\caption{\label{fig:many_lcs_and_pdf_3d}
			Concentric closed characteristics found for three-dimensional convection after removing the imperfections from the binned conditionally averaged vector field, cf.~text. The temperature PDF $f(T,z)$ is colour coded.
		}
	\end{figure}

	By smoothing the binned data through a convolution with a Gaussian kernel and projecting the vector field onto the solenoidal part (i.e., enforcing the validity of Eq.~\eqref{eq:pdfeq_3d}), we are indeed able to find the expected concentric closed curves, as exemplified in Fig.~\ref{fig:many_lcs_and_pdf_3d}.
	Here the horizontal axis corresponds to the temperature coordinate $T$ and the vertical axis to the vertical coordinate $z$ of the phase space; the background colour coding gives the temperature PDF $f(T,z)$.
	For every starting point located on the $z=0.5$-axis, the characteristics perform a closed loop in counter-clockwise direction that shows how particles on average evolve through phase space.
	By tracing its course, one is able to reconstruct the typical Rayleigh--B{\'e}nard cycle a conditional particle undergoes:
	A fluid element near the lower plate first heats up and then starts to move up towards the cold plate.
	During its upward travel it slowly cools down and then becomes much colder when it is close to the top plate before it falls down again towards the lower plate while beginning to heat up and starting what we call the RB cycle all over again.
	The cycles for fluid starting at more moderate temperatures (i.e., near $T=0.5$) show a smaller amplitude in both $T$- and $z$-direction.
	We do not find closed circles located further outwards, because the characteristics would visit areas of the phase space were no events are recorded in the numerics, and thus the vector field is undefined there.
	This usually happens near the vertical boundaries of the phase space (i.e., near the plates) where the support of the vector field becomes very narrow.
	Furthermore, we remark that the precise appearance of the closed cycles slightly varies with the parameters of the post-processing described above (e.g., the amount of smoothing); nevertheless, the description of the qualitative behaviour we are aiming at is found to be robust.
	
\begin{figure}
		\centerline{\includegraphics{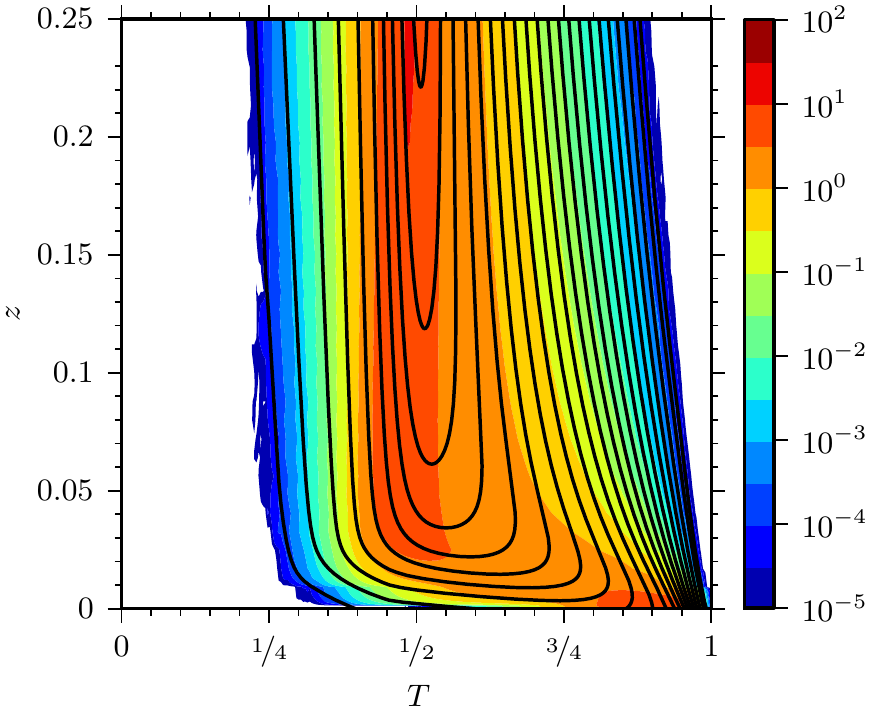}}
		\caption{\label{fig:many_lcs_and_pdf_3d_noise}
			Characteristics seeded on the $\{T>0.5, z=0.5\}$-line and followed backwards in time for three-dimensional convection. A zoom into the bottom region $0\le z\le0.25$ is shown. The divergences of the vector field have not been removed here. The temperature PDF $f(T,z)$ is colour coded.
		}
	\end{figure}

	Without the aforementioned projection to keep the vector field solenoidal in the presence of noise, the characteristics converge in some parts of the phase space.
	To study this behaviour, and also the near-wall regions, we seed characteristics on the $\{T>0.5, z=0.5\}$-line and integrate them backwards in time, as shown in Fig.~\ref{fig:many_lcs_and_pdf_3d_noise}.
	The comparison of the characteristics in the regions $T<0.5$ and $T>0.5$ immediately reveals the convergence for the non-solenoidal case, because the density of the characteristics on the right side (i.e., at later times) is higher than on the left side.
	Furthermore, the characteristics on the far right enter the boundary at $z=0$ when integrated backwards in time, or, to put it in other words, characteristics leave the boundaries very close to each other and then become less dense when following them forward in time.
	This corresponds to the fact that at $z=0$, the temperature PDF becomes a $\delta$-function located at $T=1$ due to the Dirichlet boundary condition.
	In fact, the characteristics should end precisely in this point, but this behaviour is not resolved here.
	Likewise, by symmetry considerations, the characteristics approaching the boundary from the far left side should also enter the point $\{T=1,z=0\}$, which we do not observe.
	We speculate that a numerical resolution that goes beyond what is demanded by the criterion of \citet{shishkina10njp} may be needed to capture the correct behaviour of the characteristics in this singular region of phase space.

	The cycle to which the characteristics converge in the presence of noise is shown in Fig.~\ref{fig:limit_cycle_3d_pdf}.
	This cycle is very similar to the closed curves shown in Fig.~\ref{fig:many_lcs_and_pdf_3d}; in fact, it is almost completely embedded between two adjacent closed curves.
	While the solenoidal projection helps to find the required closed curves, the conditionally averaged vector field still containing the noise gives the same general picture of the RB cycle.
	Additionally, while removing divergences, the projection may introduce unpredictable systematic errors as a side effect, especially in regions of the phase space where the support of the conditional averages and the PDF changes rapidly.
	Therefore, as we want to focus on the qualitative features of Rayleigh--B{\'e}nard convection that the RB cycle as well as the vector field represent, we will from now on only consider the cases without the solenoidal projection and use the found cycle as one generic representative of the family of closed concentric curves.
	The same applies to the convection cases discussed in Secs.~\ref{sec:dns_results_2d} and~\ref{sec:dns_results_cyl} where the characteristics also tend to converge due to imperfections induced by noise.
	
	We now come back to the discussion of the qualitative features that the conditionally averaged phase space velocity and the RB cycle describe.
	In Fig.~\ref{fig:limit_cycle_3d_pdf}, the RB cycle is shown together with the temperature PDF and in Fig.~\ref{fig:limit_cycle_3d_vel} together with the phase space velocity.
	From the first-mentioned figure, it is interesting to see that hot fluid on the RB cycle has the highest phase space speed in the range $0.25<z<0.5$, i.e. hot rising fluid is fastest in the lower half of the convection vessel, and likewise for cold fluid in the upper half due to the up-down symmetry of Rayleigh--B{\'e}nard convection.
	As a side note, from now on, whenever we describe a fluid process, the \emph{reversed} process -- interchanging hot$\leftrightarrow$cold, bottom$\leftrightarrow$top, up$\leftrightarrow$down etc.~-- is also implied.

	The temperature PDF in Fig.~\ref{fig:limit_cycle_3d_pdf} shows that the temperature distribution changes with the vertical coordinate and contracts to a $\delta$-function at the fixed temperature boundaries.
	Furthermore, one can map the shape of the distribution to the higher moments in Fig.~\ref{fig:moments_3d}, i.e.~the peaks of the standard deviation near the boundaries and the linear dependence of the skewness on the height in the bulk region.
	We note that the isocontours of the PDF do not lie tangent to the vector field or the RB cycle because the divergence of the phase space velocity, $\frac{\partial}{\partial z} \langle u_z\lvert T,z\rangle + \frac{\partial}{\partial T} \langle \Delta T\lvert T,z\rangle$, is non-vanishing.

	In Fig.~\ref{fig:limit_cycle_3d_vel}, the black and white background colour corresponds to the phase space speed, and the temperature is given by the colour of the arrows.
	This display lets us identify how fluid behaves in different parts of the phase space.
	Near the boundaries, fluid of all temperatures displays high phase space speeds, while in the bulk only fluid of \emph{intense} temperature (i.e., deviating strongly from the mean) has high speeds.
	This supports the previous finding that hot fluid on the RB cycle has its highest speed in the lower half.
	Fluid that has the mean temperature (cf.~upper-left panel of Fig.~\ref{fig:moments_3d}) is found to be at rest because no buoyancy acts on it and it does not heat up or cool down.
	As can be expected \emph{a priori}, the vector field shows that the main movement in $T$-direction, i.e.~heating and cooling, takes place near the boundaries, while the main movement in vertical direction happens in the bulk.

\subsection{Two-dimensional Convection with Periodic Horizontal Boundaries}\label{sec:dns_results_2d}
	The next case to investigate is two-dimensional Rayleigh--B{\'e}nard convection with periodic horizontal boundaries.
	The parameters are $\Ra=5\times10^8$, $\Pr=1$ and $\Gamma=4$, and the numerical scheme that is used is identical to the one from Sec.~\ref{sec:dns_results_3d}.
	Again, the two horizontal plates are no-slip walls of fixed temperature.
	The numerical resolution is $1536\times384$ equidistant grid points with $7$ grid points in the vertical direction falling into the boundary layers (cf.~\citet{shishkina10njp}), and the ensemble consists of $3891$ snapshots separated by $3.75$ free-fall time units.

	\begin{figure}
		\centerline{\includegraphics{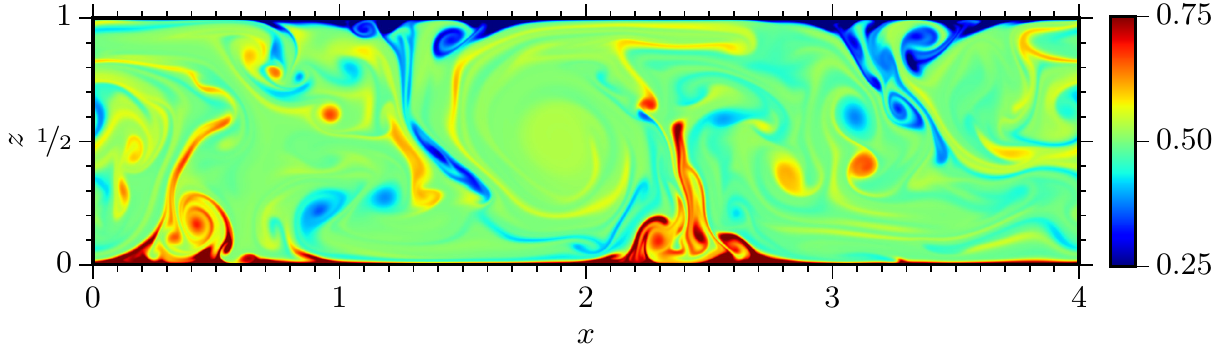}}
		\caption{\label{fig:viz_2d}
			Temperature field for two-dimensional convection with periodic horizontal boundary conditions.
			The colour scale for $T$ is shown on the right.
		}
	\end{figure}
	A snapshot of the temperature field is shown in Fig.~\ref{fig:viz_2d}, and one can see coherent structures in the form of four plume hot spots (two at the top, two at the bottom) and four convection rolls, even at this intermediate Rayleigh number.
	Also, localised round blobs of hot and cold fluid can be found.
	The statistical symmetries in this system are identical to the ones for the three-dimensional periodic case discussed before, which means the phase space becomes two-dimensional and the statistics depend on $T$ and $z$ only.

	\begin{figure}
		\centerline{\includegraphics{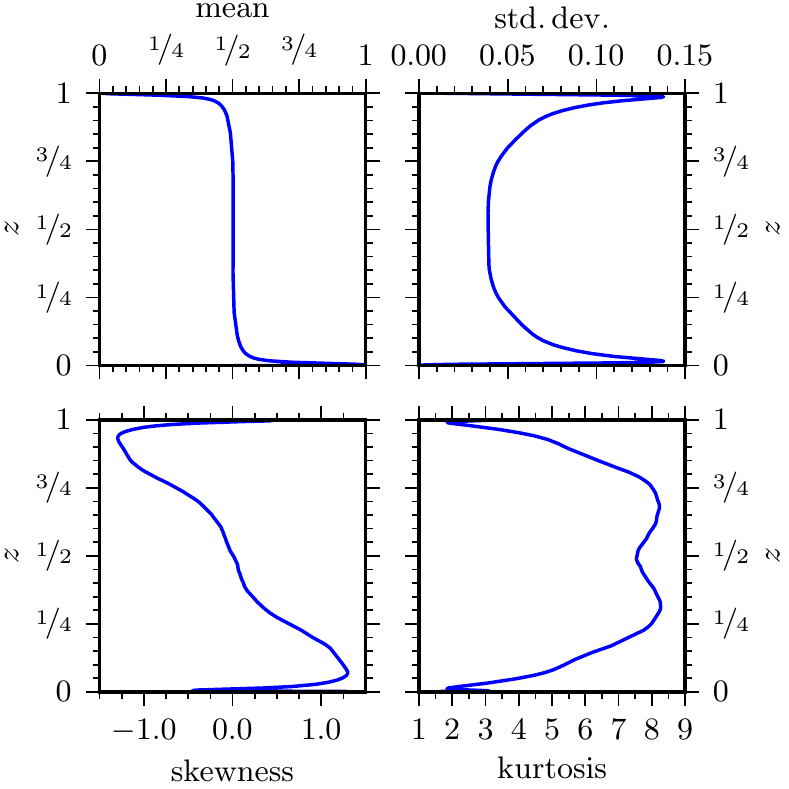}}
		\caption{\label{fig:moments_2d}
			First four standardised moments of temperature for two-dimensional convection, analogous to Fig.~\ref{fig:moments_3d}.
		}
	\end{figure}
	Figure~\ref{fig:moments_2d} shows the first four height-resolved standardised moments of temperature.
	While the mean temperature profile has the same shape as the one from three-dimensional convection (cf.~Fig.~\ref{fig:moments_3d}), the higher moments show subtle new features.
	For the three-dimensional case the moment profiles as function of the height are smoother than for the two-dimensional case.
	Especially the skewness shows transitions and is in the bulk not as linear as for the three-dimensional case.
	We link this to the coherent structures in the form of plume hot spots in the two-dimensional case because the position of the transitions in the moments corresponds to the vertical size of the plume hot spots (cf.~Fig.~\ref{fig:viz_2d}).
	In the plume hot spots, there is a re-cycling of hot fluid, which means that fluid that is hotter than the mean temperature profile is trapped near the hot bottom plate for some time instead of being advected upwards directly, cf.~\citet{sugiyama10prl}.
	This hot trapped fluid causes a temperature distribution that is near the lower plate strongly skewed towards higher temperatures.
	Above the hot spot, only a sharp jet of hot fluid remains which results in a flatter profile of the skewness.
	In three-dimensional convection the trapping mechanism of plume hot spots is missing:
	Loosely speaking, in three-dimensional convection, there is another lateral dimension into which the fluid can escape and be advected away, forming the sheet-like plumes that can also be found in Fig.~\ref{fig:viz_3d} (\citet{schmalzl04epl} and \citet{poel13jfm} also discuss differences in flow structures between two- and three-dimensional convection).
	Therefore, the entrapment seen in two-dimensional convection does not occur in three dimensions, which means that a strong mechanism that in two dimensions traps hot fluid near the bottom is missing for the three-dimensional case.

\begin{figure}
		\centerline{\includegraphics{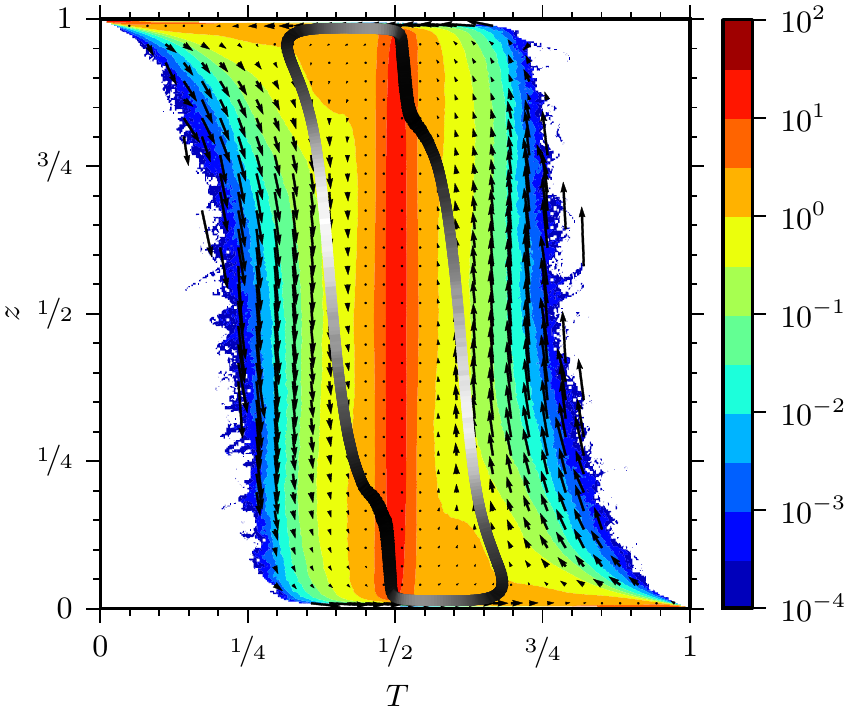}}
		\caption{\label{fig:limit_cycle_2d_pdf}
			RB cycle of the characteristics for two-dimensional convection, together with the temperature PDF.
			Illustration analogous to Fig.~\ref{fig:limit_cycle_3d_pdf}.
		}
	\end{figure}

\begin{figure}
		\centerline{\includegraphics{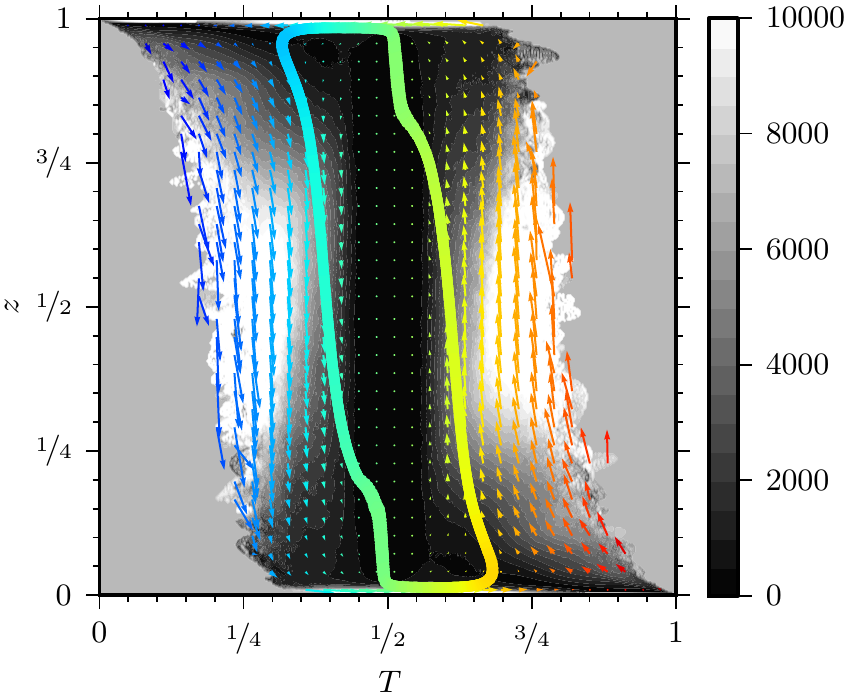}}
		\caption{\label{fig:limit_cycle_2d_vel}
			RB cycle of the characteristics for two-dimensional convection, together with the phase space speed.
			Illustration analogous to Fig.~\ref{fig:limit_cycle_3d_vel}.
		}
	\end{figure}

	When we estimate the conditional averages for two-dimensional convection from the simulation and then numerically calculate the characteristics, we again find that the dynamics in phase space resemble a closed cycle.
	This RB cycle is shown in Figs.~\ref{fig:limit_cycle_2d_pdf} and \ref{fig:limit_cycle_2d_vel}.
	While the RB cycle displays the same basic cycle of fluid heating up at the bottom, moving upwards, cooling down at the top plate and falling down again, there are some differences as compared to the three-dimensional case (Figs.~\ref{fig:limit_cycle_3d_pdf} and \ref{fig:limit_cycle_3d_vel}).

	The most striking new feature is a kink in the RB cycle in the lower left and upper right corners.
	As can be seen from the colour coding of the cycle, this is also the region of the lowest phase space speed.
	Therefore, we also link this region to the re-cycling areas discussed above because fluid trapped in the plume hot spots undergoes almost no net vertical movement and needs more time to heat up in comparison to fluid that is in direct contact with the much hotter bottom plate.
	A similar argument is discussed by \citet{poel13jfm}.

	Another difference with the three-dimensional case is that there is a bulge in the temperature PDF towards higher temperatures (around $T\approx0.6$, $z\approx0.15$).
	This bulge is due to hotter-than-average fluid that gathers near the bottom plate and is therefore compatible with the interpretation of the re-cycling fluid from above; also, this bulge gives a direct impression of the high skewness values found in this region (cf.~Fig.~\ref{fig:moments_2d}).

	Figure~\ref{fig:limit_cycle_2d_vel} shows the vector field of the phase space velocities together with its norm (coded in black and white).
	The phase space velocities are more heterogeneously distributed as compared to the three-dimensional case, e.g.~the high speeds in the bulk for intense temperatures are more pronounced (cf.~Fig.~\ref{fig:limit_cycle_3d_pdf}).
	These strong vertical movements in the bulk lead to higher phase space speeds as compared to three dimensions (see the colour scale in Figs.~\ref{fig:limit_cycle_3d_vel}~and~\ref{fig:limit_cycle_2d_vel}).
	We think this can only in part be attributed to the difference in Rayleigh numbers ($2.4\times10^7$~vs.~$5\times10^8$), but is also due to the coherent structures found in two dimensions, i.e.~plume hot spots as localised events of intense temperature.
	It is also found that in the kink region the RB cycle passes through the region of lowest phase space speed, which can be understood as the average dynamics being slowed down in the re-circulating plume hot spots.

\subsection{Three-dimensional Convection in Cylindrical Vessel}\label{sec:dns_results_cyl}
	\begin{figure}
		\centerline{\includegraphics[width=0.66\linewidth]{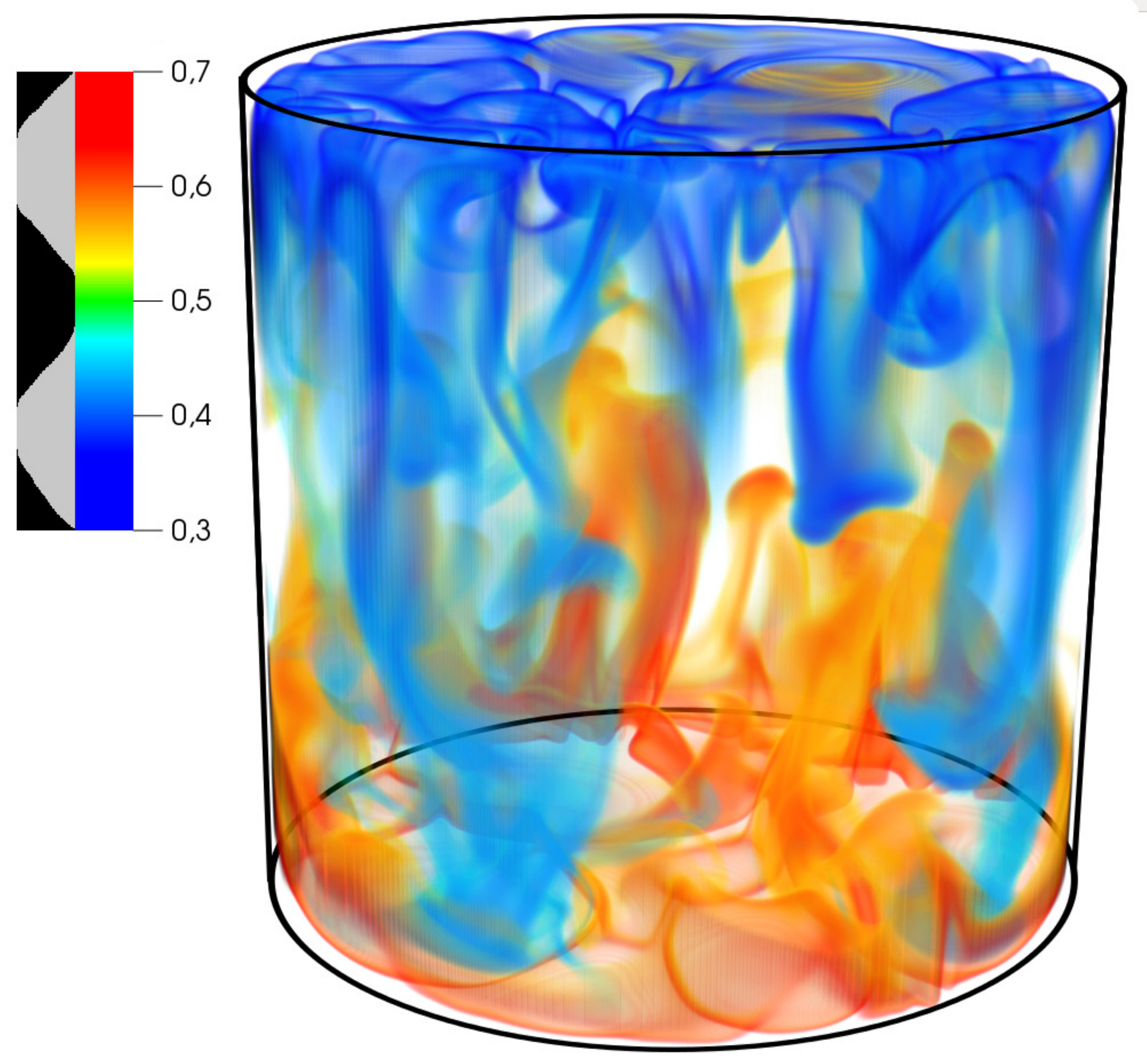}}
		\caption{\label{fig:viz_cyl}
			Snapshot of the temperature field in three-dimensional Rayleigh--B{\'e}nard convection in a $\Gamma=1$-cylinder.
			In the upper left corner, the colour and opacity scale is shown; fluid around the mean temperature is translucent.
		}
	\end{figure}
	The last Rayleigh--B{\'e}nard geometry under investigation is a closed cylindrical vessel.
	The control parameters are $\Ra=2\times10^8$, $\Pr=1$ and $\Gamma=1$ (diameter over height).
	All the walls are no-slip, and the horizontal plates are of constant temperature while the side walls are thermally insulating.
	The ensemble consists of $870$ snapshots that are obtained from direct numerical simulation using a second-order finite difference scheme on a staggered cylindrical grid \citep{verzicco03jfm} with a resolution of $N_\varphi \times N_r\times N_z=384\times192\times384$ grid points (with $\varphi$, $r$ and $z$ being the azimuthal, radial and vertical coordinate, respectively).
	The boundary layer contains $17$ grid points in the vertical direction.
	The snapshots are separated by $1$ free-fall time unit.
	Figure~\ref{fig:viz_cyl} shows a snapshot of the temperature field.

	Rayleigh--B{\'e}nard convection in a cylinder has statistical symmetries that are different from the former two cases of horizontally homogeneous convection.
	In addition to the temperature $T$ and vertical position $z$ the statistics now also depend on the radial position $r$ in the cylinder.
	Here $r=0$ corresponds to the cylinder axis and the sidewall is at $r=\nicefrac{1}{2}$.

	\begin{figure}
		\centerline{\includegraphics{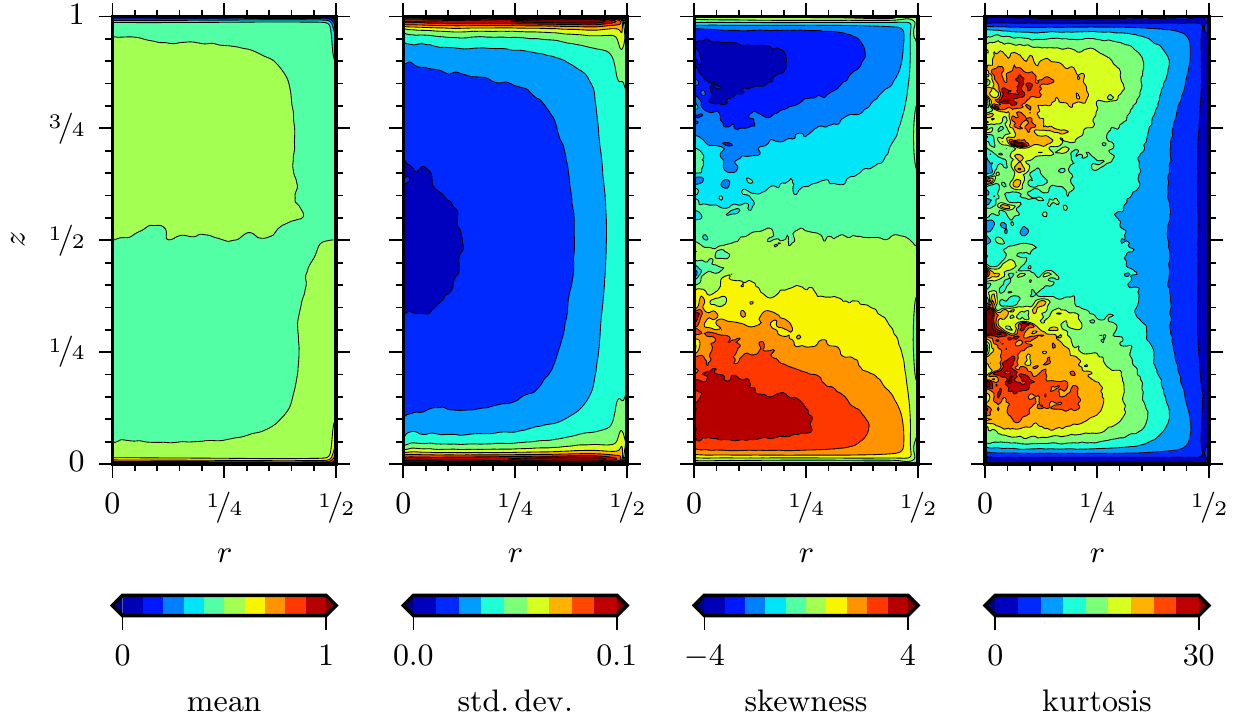}}
		\caption{\label{fig:moments_cyl}
			First four standardised $r$-$z$-resolved moments of the temperature distribution, i.e.~mean value, standard deviation, skewness and kurtosis (from left to right).
			Here the convection is in a cylinder with $\Gamma=1$.
		}
	\end{figure}
	In Fig.~\ref{fig:moments_cyl} the $r$-$z$-resolved first four standardised moments of the temperature distribution are shown.
	The horizontal and vertical axes correspond to the radial coordinate $r$ and the vertical coordinate $z$, respectively.
	The mean temperature profile is almost constant in the bulk, and only near the hot and cold plate (and to a lesser extent near the sidewalls) a deviation can be seen.
	Like in the former cases, the standard deviation of the temperature takes its highest values near the horizontal plates and falls off towards the middle of the convection cell with a local minimum at $z=0.5$.
	Also, it can be seen that this local minimum is less pronounced near the side walls, i.e.~for high $r$.
	After rising up to its maximum value at $z\approx0.1$, the skewness varies monotonically with increasing $z$.
	This has also been found for the former two cases.
	Regarding the radial dependence, the skewness falls off towards the sidewalls, indicating a less asymmetric temperature distribution there, while its highest values are found near the cylinder axis.
	Although the statistics are less converged for the kurtosis (especially near $r=0$ due to the cylindrical geometry), one can see that the highest values correspond roughly to the extrema of the skewness.
	These high values of skewness and kurtosis near the bottom wall can be attributed to hot localised plumes detaching from the hot bottom plate and piercing into the colder fluid of the bulk.
	We also note that the absolute values of skewness and kurtosis are higher than in the former two cases (Figs.~\ref{fig:moments_3d} and \ref{fig:moments_2d}), which can be understood on dimensional grounds:
	In horizontally periodic convection, we averaged over all horizontal directions and therefore averaged out the sharp maxima that can be seen in the cylindrical case (Fig.~\ref{fig:moments_cyl}) where the statistics are resolved additionally in the horizontal coordinate $r$.

	When inserting the temperature PDF $f(T,r,z)$ and the conditional averages $\langle\cdot\lvert T,r,z\rangle$ into the general framework from Sec.~\ref{sec:theory}, the PDF-defining equation \eqref{eq:pdfeq_general} becomes
	\begin{equation}\label{eq:pdfeq_cyl}
		\frac{1}{r}\frac{\partial}{\partial r}\bigl(r \langle u_r\lvert T,r,z\rangle f\bigr) + \frac{\partial}{\partial z}\bigl(\langle u_z\lvert T,r,z\rangle f\bigr) = -\frac{\partial}{\partial T} \bigl(\langle \Delta T \lvert T,r,z\rangle f\bigr),
	\end{equation}
	while the characteristics~\eqref{eq:characteristics_general} read
	\begin{equation}\label{eq:characteristics_cyl}
		\left(\begin{matrix}\dot{T} \\ \dot{r} \\ \dot{z}\end{matrix}\right) = \left(\begin{matrix}\langle\Delta T\lvert T,r,z\rangle \\ \langle u_r\lvert T,r,z\rangle \\ \langle u_z\lvert T,r,z\rangle\end{matrix}\right).
	\end{equation}
	In comparison to the former two cases, we now deal with a three-dimensional phase space where the additional dimension is related to the radial movement.
	From Eq.~\eqref{eq:characteristics_cyl} one sees that the radial coordinate $r$ of the characteristics evolves according to the conditional average of radial velocity $u_r$.
	
\begin{figure}
		\centerline{\includegraphics{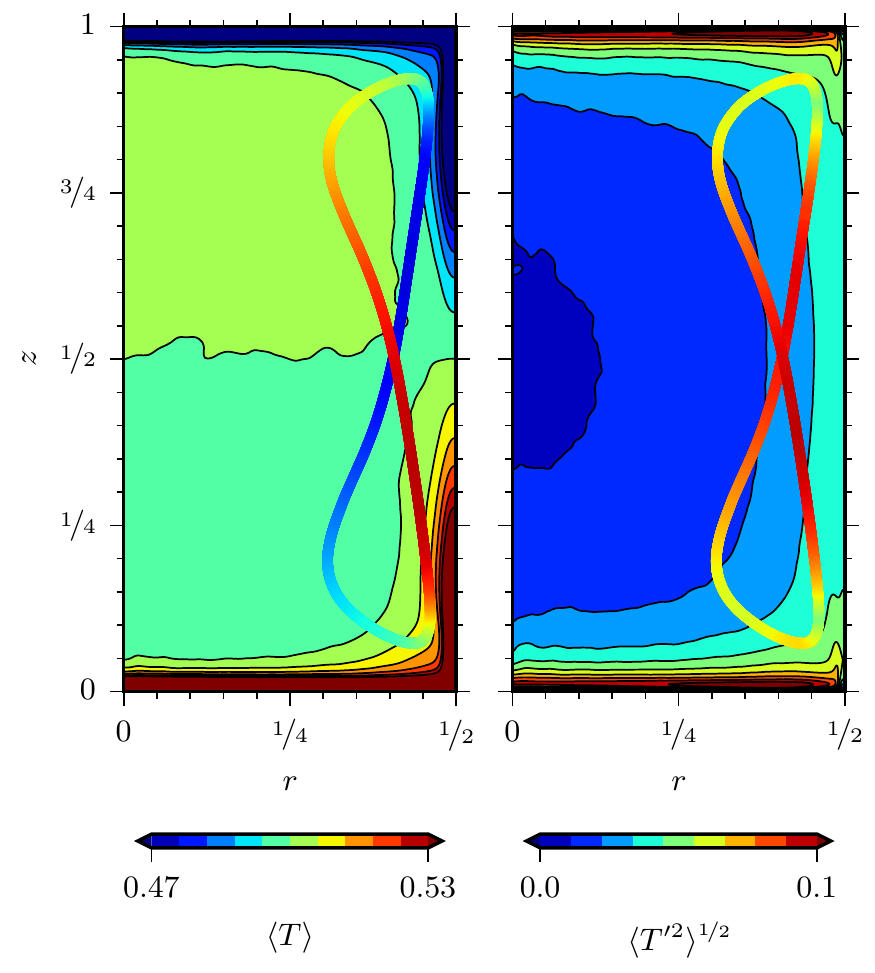}}
		\caption{\label{fig:limit_cycle_cyl}
			RB cycle of the characteristics for convection in a $\Gamma=1$-cylinder.
			Left: Temperature of the cycle (figure-eight) and $r$-$z$-resolved mean temperature colour coded.
			Right: The $r$-$z$-resolved standard deviation of temperature where $T'=T-\langle T\rangle$.
			The colour of the figure-eight-shaped cycle indicates its absolute temperature difference with respect to the background temperature.
			}
	\end{figure}

\begin{figure}
		\centerline{\includegraphics{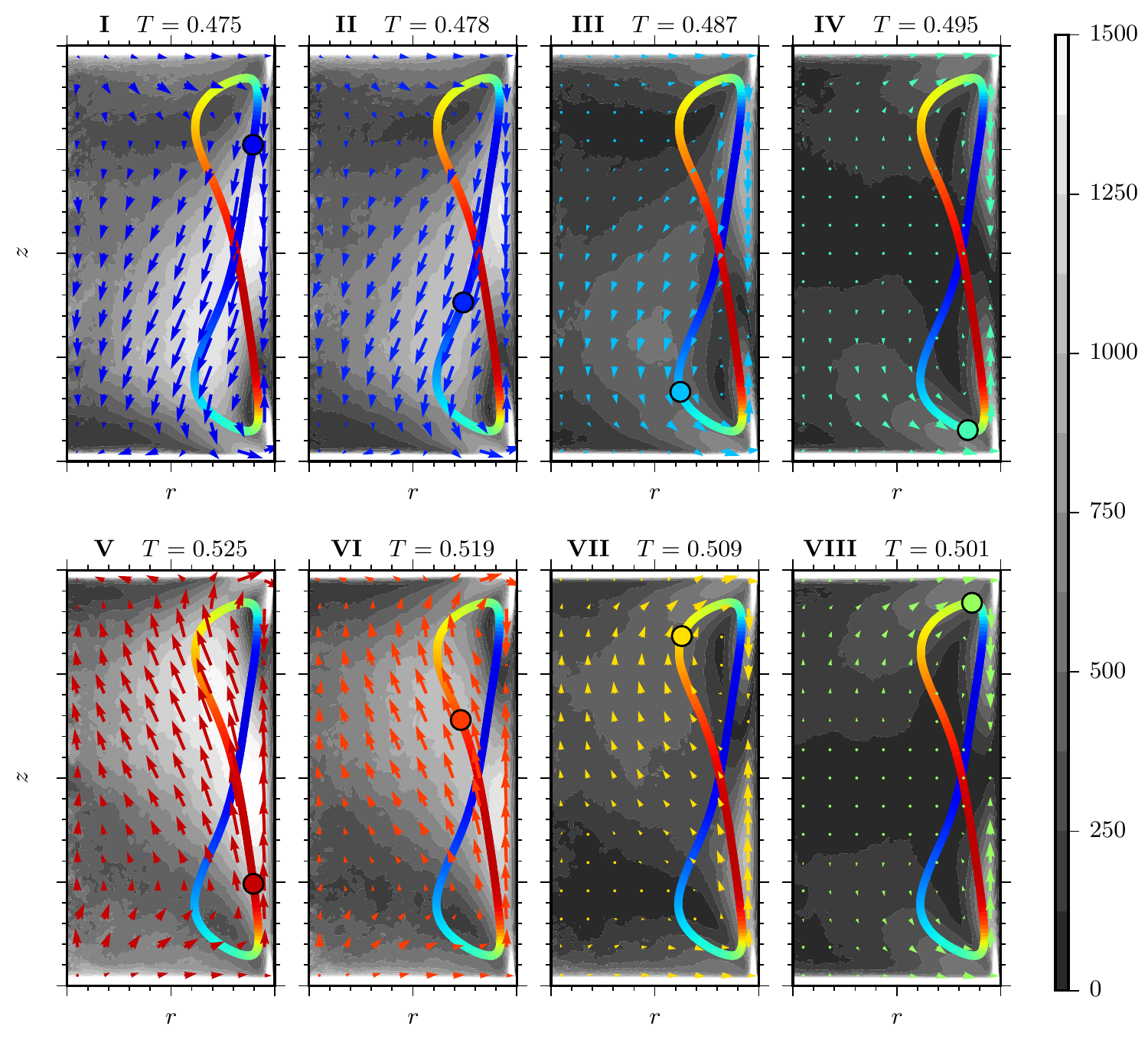}}
		\caption{\label{fig:limit_cycle_cyl_vel}
			Vector field governing the characteristics in phase space for convection in a $\Gamma=1$-cylinder, cf.~Eq.~\eqref{eq:characteristics_cyl}.
			Extents of horizontal and vertical axes as in Fig.~\ref{fig:limit_cycle_cyl}, i.e.~$0\le r\le\nicefrac{1}{2}$ and  $0\le z\le1$.
			The RB cycle is the slender figure-eight-shaped curve at the right side.
			The temperature of the vector field and the RB cycle is colour coded as in Fig.~\ref{fig:limit_cycle_cyl} (left), and the phase space speed (i.e.~norm of velocity) is coded in black and white in the background (with white being high velocity).
			The eight panels \textbf{I}--\textbf{VIII} follow a fluid parcel (circle on the cycle, with the colour showing its temperature) along the RB cycle and show a slice of the vector field of the phase space velocity in the $r$-$z$ plane at the $T$ coordinate of the fluid parcel (with $T\in\{ 0.475, 0.478, 0.487, 0.495, 0.525, 0.519, 0.509, 0.501 \}$ from left to right).
			The vector fields show the average movement in different regions of the convection cell of fluid of a particular temperature.
		}
	\end{figure}

	We now again turn to the integration of the characteristics, following Eq.~\eqref{eq:characteristics_cyl}.
	Although cylindrical convection is intrinsically different from the former two cases of horizontally periodic convection (due to three- vs.~two-dimensional phase space), we still find that the average dynamics of fluid parcels are described by a closed, twisted loop in phase space that shares common features with the former two.
	The cycle is shown in Figs.~\ref{fig:limit_cycle_cyl} and \ref{fig:limit_cycle_cyl_vel} as the slender figure-eight-shaped curve.

	In the left panel of Fig.~\ref{fig:limit_cycle_cyl}, the background shows the mean temperature (cf.~Fig.~\ref{fig:moments_cyl}) and the figure-eight-shaped curve shows a projection of the RB cycle into the $r$-$z$-plane.
	The third coordinate of the RB cycle, the temperature $T$, is colour coded.
	The temperature scale corresponds to the minimal and maximal temperature ($T=0.47$ and $T=0.53$) the RB cycle takes.
	When tracing the cycle, one can again identify the Rayleigh--B{\'e}nard cycle of the horizontally periodic convection cases, superposed with an additional inwards and outwards motion:
	Starting with fluid of mean temperature that is quickly heating up at the bottom, it then begins to rise up and move inwards into the bulk until $z\approx0.8$ and $r\approx0.3$, where it goes outwards and starts to cool down.
	At the maximal $z$, the fluid cools down quickly and then falls towards the lower plate while moving inwards, thus starting the RB cycle all over again.
	Additionally, one can see that the hot fluid rising from the lower plate steadily cools down when it crosses the bulk of almost uniform temperature; this is related to the monotonically decreasing skewness of temperature that can be seen in Fig.~\ref{fig:moments_cyl}.

	The difference of the temperature of the RB cycle and the background temperature (cf.~colour coding of these two in Fig.~\ref{fig:limit_cycle_cyl}~(left)) shows that the regions where the temperature of the RB cycle deviates most from the mean background temperature are the regions of high buoyancy and correspond to the regions of main vertical movement in the bulk.
	To elaborate on this, the right panel of Fig.~\ref{fig:limit_cycle_cyl} shows the standard deviation of the temperature field in the background, and the colour coding of the cycle shows the absolute deviation of its temperature coordinate $T$ from the mean temperature.
	The deviation of the temperature of a fluid particle on the cycle from the surrounding mean temperature determines its mean buoyancy, so the right panel tells us how strong the buoyancy acts; the highest values for hot rising fluid are found in the lower half of the convection cell.
	In comparison, the mean deviation of fluid from the mean temperature profile (shown in the background as the standard deviation of temperature) is much weaker.
	To summarise, from the left panel of Fig.~\ref{fig:limit_cycle_cyl} one can see in which direction the buoyancy acts, while the right panel shows its strength.

	The vector field of the characteristics is shown in Fig.~\ref{fig:limit_cycle_cyl_vel}.
	Due to the difficulty to display a vector field in three-dimensional phase space, we show slices of the phase space velocity in the $r$-$z$ plane at different $T$.
	The panels \textbf{I}--\textbf{VIII} show one cycle of a fluid parcel travelling along the RB cycle, with its temperature colour coded as in Fig.~\ref{fig:limit_cycle_cyl} (left).
	Additionally to the RB cycle described above, Fig.~\ref{fig:limit_cycle_cyl_vel} also reveals the average behaviour of fluid in different parts of the convection cell, conditioned on its temperature.
	The arrows show an $r$-$z$ slice of the vector field of the characteristics \eqref{eq:characteristics_cyl} at the $T$ coordinate of the cycle.
	Also, the black and white background colour indicates the phase space speed, i.e.~how fast a fluid parcel travels through phase space (with white being the fastest movement).
	The arrows indicate the mean movement of fluid of a given temperature in different regions of the convection cell.
	
	Panel \textbf{I} shows that cold fluid (here, $T=0.475$) has the highest speeds in the bulk and near the sidewall.
	Near the cold top plate, cold fluid is mainly transported towards the outer wall.
	For $z<0.75$, the direction of movement is slightly tilted towards the cylinder axis.
	Cold fluid that falls down along the sidewall is deflected towards the inner cylinder at around $z\approx0.25$ due to a corner flow that propels cold fluid upwards along the side wall; notice that at $r\approx0.45$ and $z\approx0.25$, up- and downwelling cold fluid collides.
	This feature can be understood as cold plumes that are formed at the upper plate and are swept towards the sidewalls.
	The plumes then fall down along the sidewall until they hit the hot fluid at the bottom plate where they are directed inwards.
	These cold plumes that fall down along the sidewall can also be seen in Fig.~\ref{fig:viz_cyl}.

	The fluid from panel \textbf{III} is less cold ($T=0.487$) and has overall lower speeds, but still shows the same features as in panel \textbf{I}, e.g.~cold fluid is swept along the upper plate outwards and falls down along the side wall until it hits the upwelling corner flow.
	The vector fields for fluid of mean temperature ($T\approx0.5$, panel \textbf{IV} and \textbf{VIII}) are symmetric around $z=0.5$ and show an almost vanishing velocity in the bulk.
	Near the horizontal plates, fluid is swept towards the sidewalls and from there vertically towards the $z=0.5$-line.

	The rest of the panels complete one run of the RB cycle and due to the up-down symmetry contain the same information already described.
	Still, from all eight panels it can be seen that fluid for all temperatures has high phase space speed near the sidewalls, which can be attributed to plumes that are guided along the outer walls of the cylinder, and also has high speeds near the bottom and top plate which is due to the vigorous temperature contrast between fluid and horizontal plates that leads to high speeds regarding the $T$ coordinate.
	Also, we have to stress here that the corner flows show up due to the investigation being conditioned on the temperature and not due to some large-scale structure that may be present in the fields; this structure is lost in the azimuthal averaging process when obtaining the vector field from Eq.~\eqref{eq:characteristics_cyl}.
	Therefore, in our analysis the corner flows are statistical structures that are not necessarily related to structures in the flow.

\section{Summary}\label{sec:discussion}
	In this paper, we analysed the turbulent flow in Rayleigh--B{\'e}nard convection on the basis of statistical quantities like the temperature PDF and conditionally averaged fields.
	We derived that the mean path a fluid particle takes through phase space (spanned by temperature and spatial coordinates) is defined by so-called \emph{characteristics}, i.e.~trajectories in phase space that follow the conditionally averaged vector field composed of heat diffusion and fluid velocities.
	Thereby, we could characterise the dynamics and flow features that occur in turbulent convection cells from a statistical point of view, i.e.~from averaged quantities like the temperature distribution and its moments as well as statistics conditioned on temperature and spatial position.

	By estimating the aforementioned vector fields for three different Rayleigh--B{\'e}nard geometries while utilising their symmetries and then integrating the characteristics we described the mean dynamics that fluid particles undergo, i.e.~we could describe how fluid of different temperatures behaves in different regions of the convection volume.
	We also distinguished regions of high and low transport through phase space.
	For all geometries there are high phase space speeds for intense temperatures in the bulk (which we attribute to localised events of intense temperatures and high speeds, i.e.~plumes) as well as high speeds near the horizontal plates for all temperatures, while for the case of cylindrical convection the phase space speed also takes high values near the wall of the cylinder.
	This we interpret as plumes that are directed along the insulating sidewalls.
	In the conditionally averaged vector field of the cylinder, we could furthermore identify corner flows near the sidewalls for fluid of different temperatures.
	Cold fluid experiences a corner flow near the bottom plate while showing no corner-flow near the upper plate and vice-versa.
	Additionally, we described the higher moments of the temperature distributions, where we could link features of the moments to coherent structures that appear in turbulent flows.

	When we then obtained the characteristics by integrating trajectories through the conditionally averaged vector field,  we found that for all different convection setups, the characteristics form closed cycles in phase space.
	These cycles reconstruct the typical Rayleigh--B{\'e}nard cycle a fluid particle undergoes on average, i.e.~fluid is heated up at the bottom and rises upwards while slightly cooling down until it hits the upper plate, where it cools down fast and falls down to the lower plate while slightly heating up, thus starting the cycle all over again.
	In the cylindrical case, where there is another phase space dimension corresponding to horizontal movement, the fluid shows an additional inwards and outwards movement while following the RB cycle.
	The method thus allows to further pin-point and quantify the differences and similarities between Rayleigh--B{\'e}nard convection in two- and three-dimensional periodic boxes and in three-dimensional convection in a cylindrical cell.

	For so-called homogeneous Rayleigh--B{\'e}nard convection \citep{lohse03prl, calzavarini05pof} -- thermal convection with horizontally periodic boundary conditions as well as periodic boundary conditions in vertical direction together with an imposed temperature gradient driving the flow -- we would expect quite different behaviour as such a system does not have boundary layers, but represents pure bulk turbulence.
	Furthermore, the statistics do not depend on the spatial coordinates and the phase space becomes one-dimensional, which is fundamentally different from the three cases considered in the present paper.
	The analysis of homogeneous Rayleigh--B{\'e}nard convection would actually be more in line with the work by \citet{yakhot89prl} and \citet{ching93prl}, who analyse with the help of conditional averages the PDF of experimental time series obtained from temperature probe measurements.
	In these works, in a sense the phase space is also one-dimensional due to the lack of any spatial coordinate.
	Thus, while their ansatz can be used to e.g.~describe the transition from soft to hard turbulence, no information about the dynamics and spatial structures can be obtained from the statistics, contrary to the method we proposed here.

	The authors thank Prof. Stephen~B.~Pope for his insightful remarks regarding the interpretation of our results and the referees for their constructive input.
	JL acknowledges fruitful discussions with and valuable advice by Oliver Kamps, Anton Daitche and Theodore Drivas as well as funding by the DFG (Deutsche Forschungsgemeinschaft) under grant FR~1003/10-1.
	MW acknowledges support from the DFG under project WI~3544/2-1 and WI~3544/3-1.
	RJAMS was supported by the \emph{Fellowships for Young Energy Scientists} (YES!) of FOM.
	Parts of the computations have been performed on PROVIDE and Huygens (SURFsara).

\bibliographystyle{jfm}
\bibliography{ref_abbrv}

\begin{thebibliography}{44}
\expandafter\ifx\csname natexlab\endcsname\relax\def\natexlab#1{#1}\fi

\bibitem[Ahlers {\em et~al.\/}(2009)Ahlers, Grossmann \& Lohse]{ahlers09rmp}
{\sc Ahlers, G., Grossmann, S. \& Lohse, D.} 2009 Heat transfer and large scale
  dynamics in turbulent {R}ayleigh--{B}{\'e}nard convection. {\em Rev. Mod.
  Phys.\/} {\bf 81}~(2), 503--537.

\bibitem[Ahlers {\em et~al.\/}(2012)Ahlers, He, Funfschilling \&
  Bodenschatz]{ahlers12njp}
{\sc Ahlers, G., He, X., Funfschilling, D. \& Bodenschatz, E.} 2012 Heat
  transport by turbulent {R}ayleigh--{B}{\'e}nard convection for
  $\mathrm{Pr} \simeq 0.8$ and $3\times10^{12}\lesssim
  \mathrm{Ra}\lesssim10^{15}$: aspect ratio {$\Gamma = 0.5$}. {\em New J.
  Phys.\/} {\bf 14}~(10), 103012.

\bibitem[Angot {\em et~al.\/}(1999)Angot, Bruneau \& Fabrie]{angot99num}
{\sc Angot, P., Bruneau, C.-H. \& Fabrie, P.} 1999 A penalization method to
  take into account obstacles in incompressible viscous flows. {\em Numer.
  Math.\/} {\bf 81}~(4), 497--520.

\bibitem[Bailon-Cuba {\em et~al.\/}(2010)Bailon-Cuba, Emran \&
  Schumacher]{bailoncuba10jfm}
{\sc Bailon-Cuba, J., Emran, M.~S. \& Schumacher, J.} 2010 Aspect ratio
  dependence of heat transfer and large-scale flow in turbulent convection.
  {\em J. Fluid Mech.\/} {\bf 655}, 152--173.

\bibitem[Boussinesq(1903)]{boussinesq03book}
{\sc Boussinesq, J.} 1903 {\em Th{\'e}orie Analytique de la Chaleur\/}. Paris:
  Gauthier-Villars.

\bibitem[Calza\-varini {\em et~al.\/}(2005)Calza\-varini, Lohse, Toschi \&
  Tripiccione]{calzavarini05pof}
{\sc Calza\-varini, E., Lohse, D., Toschi, F. \& Tripiccione, R.} 2005
  {R}ayleigh and {P}randtl number scaling in the bulk of
  {R}ayleigh--{B}{\'e}nard turbulence. {\em Phys. Fluids\/} {\bf 17}~(5),
  055107.

\bibitem[Chill{\`a} \& Schumacher(2012)]{chilla12epj}
{\sc Chill{\`a}, F. \& Schumacher, J.} 2012 New perspectives in turbulent
  {R}ayleigh--{B}{\'e}nard convection. {\em Eur. J. Phys. E\/} {\bf 35}, 1--25.

\bibitem[Ching(1993)]{ching93prl}
{\sc Ching, E. S.~C.} 1993 Probability densities of turbulent temperature
  fluctuations. {\em Phys. Rev. Lett.\/} {\bf 70}~(3), 283--286.

\bibitem[Ching {\em et~al.\/}(2004)Ching, Guo, Shang, Tong \& Xia]{ching04prl}
{\sc Ching, E. S.~C., Guo, H., Shang, X.-D., Tong, P. \& Xia, K.-Q.} 2004
  Extraction of plumes in turbulent thermal convection. {\em Phys. Rev.
  Lett.\/} {\bf 93}, 124501.

\bibitem[Courant \& Hilbert(1962)]{courant62book}
{\sc Courant, R. \& Hilbert, D.} 1962 {\em Methods of Mathematical Physics
  Volume II\/}. Wiley-Interscience.

\bibitem[Friedrich {\em et~al.\/}(2012)Friedrich, Daitche, Kamps, L{\"u}lff,
  Vo{\ss}kuhle \& Wilczek]{friedrich12crp}
{\sc Friedrich, R., Daitche, A., Kamps, O., L{\"u}lff, J., Vo{\ss}kuhle, M. \&
  Wilczek, M.} 2012 The {L}undgren--{M}onin--{N}ovikov hierarchy: {K}inetic
  equations for turbulence. {\em Compt. Rend. Phys.\/} {\bf 13}~(9--10),
  929--953.

\bibitem[Gasteuil {\em et~al.\/}(2007)Gasteuil, Shew, Gibert, Chill{\`a},
  Castaing \& Pinton]{gasteuil07prl}
{\sc Gasteuil, Y., Shew, W.~L., Gibert, M., Chill{\`a}, F., Castaing, B. \&
  Pinton, J.-F.} 2007 {L}agrangian temperature, velocity, and local heat flux
  measurement in {R}ayleigh--{B}{\'e}nard convection. {\em Phys. Rev. Lett.\/}
  {\bf 99}~(23), 234302.

\bibitem[Grossmann \& Lohse(2000)]{grossmann00jfm}
{\sc Grossmann, S. \& Lohse, D.} 2000 Scaling in thermal convection: a unifying
  theory. {\em J. Fluid Mech.\/} {\bf 407}, 27--56.

\bibitem[Grossmann \& Lohse(2001)]{grossmann01prl}
{\sc Grossmann, S. \& Lohse, D.} 2001 Thermal convection for large {P}randtl
  numbers. {\em Phys. Rev. Lett.\/} {\bf 86}~(15), 3316--3319.

\bibitem[Grossmann \& Lohse(2011)]{grossmann11pof}
{\sc Grossmann, S. \& Lohse, D.} 2011 Multiple scaling in the ultimate regime
  of thermal convection. {\em Phys. Fluids\/} {\bf 23}~(4), 045108.

\bibitem[Grossmann \& Lohse(2012)]{grossmann12pof}
{\sc Grossmann, S. \& Lohse, D.} 2012 Logarithmic temperature profiles in the
  ultimate regime of thermal convection. {\em Phys. Fluids\/} {\bf 24}~(12),
  125103.

\bibitem[Keetels {\em et~al.\/}(2007)Keetels, D'Ortona, Kramer, Clercx,
  Schneider \& van Heijst]{keetels07jcp}
{\sc Keetels, G.~H., D'Ortona, U., Kramer, W., Clercx, H. J.~H., Schneider, K.
  \& van Heijst, G. J.~F.} 2007 {F}ourier spectral and wavelet solvers for the
  incompressible {N}avier-{S}tokes equations with volume-penalization:
  Convergence of a dipole-wall collision. {\em J. Comput. Phys.\/} {\bf
  227}~(2), 919--945.

\bibitem[Lohse \& Toschi(2003)]{lohse03prl}
{\sc Lohse, D. \& Toschi, F.} 2003 Ultimate state of thermal convection. {\em
  Phys. Rev. Lett.\/} {\bf 90}, 034502.

\bibitem[Lohse \& Xia(2010)]{lohse10rfm}
{\sc Lohse, D. \& Xia, K.-Q.} 2010 Small-scale properties of turbulent
  {R}ayleigh--{B}{\'e}nard convection. {\em Annu. Rev. Fluid Mech.\/} {\bf
  42}~(1), 335--364.

\bibitem[L{\"u}lff {\em et~al.\/}(2011)L{\"u}lff, Wilczek \&
  Friedrich]{luelff11njp}
{\sc L{\"u}lff, J., Wilczek, M. \& Friedrich, R.} 2011 Temperature statistics
  in turbulent {R}ayleigh--{B}{\'e}nard convection. {\em New J. Phys.\/} {\bf
  13}~(1), 015002.

\bibitem[Lundgren(1967)]{lundgren67pof}
{\sc Lundgren, T.~S.} 1967 Distribution functions in the statistical theory of
  turbulence. {\em Phys. Fluids\/} {\bf 10}~(5), 969--975.

\bibitem[Monin(1967)]{monin67pmm}
{\sc Monin, A.~S.} 1967 Equations of turbulent motion. {\em Prikl. Mat.
  Mekh.\/} {\bf 31}~(6), 1057--1068.

\bibitem[Novikov(1968)]{novikov68sdp}
{\sc Novikov, E.~A.} 1968 Kinetic equations for a vortex field. {\em Sov.
  Phys.-Dokl.\/} {\bf 12}~(11), 1006--1008.

\bibitem[Oberbeck(1879)]{oberbeck79apc}
{\sc Oberbeck, A.} 1879 {\"U}ber die {W}{\"a}rmeleitung der {F}l{\"u}ssigkeiten
  bei {B}er{\"u}cksichtigung der {S}tr{\"o}mungen infolge von
  {T}emperaturdifferenzen. {\em Ann. Phys. Chem.\/} {\bf 7}, 271--292.

\bibitem[Petschel {\em et~al.\/}(2013)Petschel, Stellmach, Wilczek, L{\"u}lff
  \& Hansen]{petschel13prl}
{\sc Petschel, K., Stellmach, S., Wilczek, M., L{\"u}lff, J. \& Hansen, U.}
  2013 Dissipation layers in {R}ayleigh--{B}{\'e}nard convection: A unifying
  view. {\em Phys. Rev. Lett.\/} {\bf 110}, 114502.

\bibitem[Petschel {\em et~al.\/}(2011)Petschel, Wilczek, Breuer, Friedrich \&
  Hansen]{petschel11pre}
{\sc Petschel, K., Wilczek, M., Breuer, M., Friedrich, R. \& Hansen, U.} 2011
  Statistical analysis of global wind dynamics in vigorous
  {R}ayleigh--{B}{\'e}nard convection. {\em Phys. Rev. E\/} {\bf 84}~(2),
  026309.

\bibitem[van~der Poel {\em et~al.\/}(2014)van~der Poel, Ostilla-M{\'o}nico,
  Verzicco \& Lohse]{poel14pre}
{\sc van~der Poel, E.~P., Ostilla-M{\'o}nico, R., Verzicco, R. \& Lohse, D.}
  2014 Effect of velocity boundary conditions on the heat transfer and flow
  topology in two-dimensional {R}ayleigh--{B}{\'e}nard convection. {\em Phys.
  Rev. E\/} {\bf 90}, 013017.

\bibitem[van~der Poel {\em et~al.\/}(2011)van~der Poel, Stevens \&
  Lohse]{poel11pre}
{\sc van~der Poel, E.~P., Stevens, R. J. A.~M. \& Lohse, D.} 2011 Connecting
  flow structures and heat flux in turbulent {R}ayleigh--{B}{\'e}nard
  convection. {\em Phys. Rev. E\/} {\bf 84}~(4), 045303.

\bibitem[van~der Poel {\em et~al.\/}(2013)van~der Poel, Stevens \&
  Lohse]{poel13jfm}
{\sc van~der Poel, E.~P., Stevens, R. J. A.~M. \& Lohse, D.} 2013 Comparison
  between two- and three-dimensional {{Rayleigh--B\'enard}} convection. {\em J.
  Fluid Mech.\/} {\bf 736}, 177--194.

\bibitem[Pope(1984)]{pope84iaa}
{\sc Pope, S.~B.} 1984 Calculations of a plane turbulent jet. {\em AIAA
  Journal\/} {\bf 22}~(7), 896--904.

\bibitem[Pope(1985)]{pope85pec}
{\sc Pope, S.~B.} 1985 {PDF} methods for turbulent reactive flows. {\em Prog.
  Energy Combust. Sci.\/} {\bf 11}, 119--192.

\bibitem[Pope(2000)]{pope00book}
{\sc Pope, S.~B.} 2000 {\em Turbulent Flows\/}. Cambridge, England: Cambridge
  University Press.

\bibitem[Sarra(2003)]{sarra03joma}
{\sc Sarra, S.~A.} 2003 The method of characteristics with applications to
  conservation laws. {\em J. Onl. Math. Appl.\/} {\bf 3}, 1--6.

\bibitem[Schmalzl {\em et~al.\/}(2004)Schmalzl, Breuer, Wessling \&
  Hansen]{schmalzl04epl}
{\sc Schmalzl, J., Breuer, M., Wessling, S. \& Hansen, U.} 2004 On the validity
  of two-dimensional numerical approaches to time-dependent thermal convection.
  {\em Europhys. Lett.\/} {\bf 67}~(3), 390--396.

\bibitem[Schneider(2005)]{schneider05caf}
{\sc Schneider, K.} 2005 Numerical simulation of the transient flow behaviour
  in chemical reactors using a penalisation method. {\em Comput. Fluids\/} {\bf
  34}~(10), 1223--1238.

\bibitem[Schumacher(2009)]{schumacher09pre}
{\sc Schumacher, J.} 2009 {L}agrangian studies in convective turbulence. {\em
  Phys. Rev. E\/} {\bf 79}~(5), 056301.

\bibitem[Shang {\em et~al.\/}(2008)Shang, Tong \& Xia]{shang08prl}
{\sc Shang, X.-D., Tong, P. \& Xia, K.-Q.} 2008 Scaling of the local convective
  heat flux in turbulent {R}ayleigh--{B}{\'e}nard convection. {\em Phys. Rev.
  Lett.\/} {\bf 100}, 244503.

\bibitem[Shishkina {\em et~al.\/}(2010)Shishkina, Stevens, Grossmann \&
  Lohse]{shishkina10njp}
{\sc Shishkina, O., Stevens, R. J. A.~M., Grossmann, S. \& Lohse, D.} 2010
  Boundary layer structure in turbulent thermal convection and its consequences
  for the required numerical resolution. {\em New J. Phys.\/} {\bf 12}~(7),
  075022.

\bibitem[Stevens {\em et~al.\/}(2013)Stevens, van~der Poel, Grossmann \&
  Lohse]{stevens13jfm}
{\sc Stevens, R. J. A.~M., van~der Poel, E.~P., Grossmann, S. \& Lohse, D.}
  2013 The unifying theory of scaling in thermal convection: The updated
  prefactors. {\em J. Fluid Mech.\/} {\bf 730}, 295--308.

\bibitem[Sugiyama {\em et~al.\/}(2010)Sugiyama, Ni, Stevens, Chan, Zhou, Xi,
  Sun, Grossmann, Xia \& Lohse]{sugiyama10prl}
{\sc Sugiyama, K., Ni, R., Stevens, R. J. A.~M., Chan, T.~S., Zhou, S.-Q., Xi,
  H.-D., Sun, C., Grossmann, S., Xia, K.-Q. \& Lohse, D.} 2010 Flow reversals
  in thermally driven turbulence. {\em Phys. Rev. Lett.\/} {\bf 105}~(3),
  034503.

\bibitem[Verzicco \& Camussi(2003)]{verzicco03jfm}
{\sc Verzicco, R. \& Camussi, R.} 2003 Numerical experiments on strongly
  turbulent thermal convection in a slender cylindrical cell. {\em J. Fluid
  Mech.\/} {\bf 477}, 19--49.

\bibitem[Wilczek {\em et~al.\/}(2011)Wilczek, Daitche \&
  Friedrich]{wilczek11jfm}
{\sc Wilczek, M., Daitche, A. \& Friedrich, R.} 2011 On the velocity
  distribution in homogeneous isotropic turbulence: {c}orrelations and
  deviations from {G}aussianity. {\em J. Fluid Mech.\/} {\bf 676}, 191--217.

\bibitem[Wilczek \& Friedrich(2009)]{wilczek09pre}
{\sc Wilczek, M. \& Friedrich, R.} 2009 Dynamical origins for non-{G}aussian
  vorticity distributions in turbulent flows. {\em Phys. Rev. E\/} {\bf
  80}~(1), 016316.

\bibitem[Yakhot(1989)]{yakhot89prl}
{\sc Yakhot, V.} 1989 Probability distributions in high-{R}ayleigh number
  {B}{\'e}nard convection. {\em Phys. Rev. Lett.\/} {\bf 63}~(18), 1965--1967.

\end{thebibliography}

\end{document}